\documentclass{article}
\usepackage{arxiv}
\usepackage{makecell}
\usepackage{authblk}
\usepackage[utf8]{inputenc} % allow utf-8 input
\usepackage[T1]{fontenc}    % use 8-bit T1 fonts
\usepackage{hyperref}       % hyperlinks
\usepackage{url}            % simple URL typesetting
\usepackage{booktabs}       % professional-quality tables
\usepackage{amsfonts}       % blackboard math symbols
\usepackage{nicefrac}       % compact symbols for 1/2, etc.
\usepackage{microtype}      % microtypography 
\usepackage{lipsum}
\usepackage{amsthm}
\usepackage{amsmath}
\usepackage{amssymb}
\usepackage{rotating,  graphicx}
\usepackage{cancel}					% Cancel something to something different (Arrow through equation)
\usepackage{changepage}		% used to insert whitespace left of a paragraph
\usepackage{subcaption}		% Include subfigures
\usepackage{pdfpages}		% Include external pdfs
\usepackage{wrapfig}
\usepackage{tikz}		% Create circled numbers

    \makeatletter
\def\@fnsymbol#1{\ensuremath{\ifcase#1\or \dagger\or \ddagger\or
   \mathsection\or \mathparagraph\or \|\or **\or \dagger\dagger
   \or \ddagger\ddagger \else\@ctrerr\fi}}

\usepackage{framed}

\usepackage[numbers,sort&compress]{natbib}
 
\usepackage{verbatim} % comment a block

\DeclareGraphicsExtensions{.eps}
\title{In silico modeling for personalized stenting in aortic coarctation}
\author[1,2,*]{Dandan Ma}
\author[2,3,*]{Yong Wang}
\author[1,2]{Mueed Azhar}
\author[4]{Ansgar Adler}
\author[2,5,a]{Michael Steinmetz}
\author[1,2,4,6,7,a]{Martin Uecker }

\affil[1]{\footnotesize Institute of Diagnostic and Interventional Radiology, University Medical Center G\"ottingen, 37075 G\"ottingen, Germany}
\affil[2]{\footnotesize DZHK (German Center for Cardiovascular Research), Partner Site Göttingen, 37075 G\"ottingen, Germany}
\affil[3]{\footnotesize Laboratory for Fluid Physics, Pattern Formation and Biocomplexity, Max Planck Institute for Dynamics and Self-Organization, 37077 G\"ottingen, Germany}  
\affil[4]{\footnotesize Institute of Biomedical Imaging, Graz University of Technology, 8010 Graz, Austria}
\affil[5]{\footnotesize  Department of Pediatric Cardiology and Intensive Care Medicine, University Medical Center Göttingen, 37075 Göttingen, Germany}
\affil[6]{\footnotesize Cluster of Excellence ``Multiscale Bioimaging: from Molecular Machines to Networks of Excitable Cells'' (MBExC), University of Göttingen, Germany}
\affil[7]{\footnotesize BioTechMed-Graz, Austria}

\affil[*]{\footnotesize Corresponding authors: DDM, dandan.ma@stud.uni-goettingen.de; YW, yong.wang@ds.mpg.de}

\affil[a]{\footnotesize MS and MU contributed equally.}

\date{}

\begin{document}
\maketitle

\begin{abstract}
Stent intervention is a recommended therapy to reduce the pressure gradient and
restore blood flow for patients with coarctation of the aorta (CoA). In
this work, we developed a framework for personalized stent intervention
in CoA using in silico modeling, combining computational fluid dynamics
(CFD) and image-based prediction of the geometry of the aorta after
stent intervention. Firstly, the blood flow in the aorta, whose
geometry was reconstructed from magnetic resonance imaging (MRI) data,
was numerically modeled using the lattice Boltzmann method (LBM). Both
large eddy simulation (LES) and direct numerical simulation (DNS) were
considered to adequately resolve the turbulent hemodynamics, with boundary
conditions extracted from phase-contrast flow MRI. By comparing the results
from CFD and 4D-Flow MRI in 3D-printed flow phantoms, we concluded that the
LBM based LES is capable of obtaining accurate aortic flow with acceptable
computational cost. In silico stent implantation for a patient with CoA
was then performed by predicting the deformed geometry after stent
intervention and predicting the blood flow. By evaluating the pressure
drop and maximum wall shear stress, an optimal stent can be selected.
\end{abstract}


\section{Introduction}
Coarctation of the aorta (CoA) refers to a local narrowing of the aortic arch. It makes up 6 - 8$\%$ of all congenital heart diseases \cite{Rafieianzab2021}, and is often associated with other cardiovascular diseases, such as aortic arch hypoplasia, subaortic stenosis, ventricular and atrial septal defects \cite{Hoffman2002,Reller2008,Alkashkari2019}. The coarctation leads to high blood pressure and thus heart damage. Stent intervention, which is generally performed based on clinical experience, is a recommended therapy to reduce the pressure gradient and restore blood flow. 

With increasing computational power, in silico modeling is emerging as a
promising tool to help clinicians with intervention planning and to evaluate the outcome of therapies, such as stenting for intracranial aneurysm \cite{Zhong2016,Berg2018}, abdominal aortic aneurysm \cite{ Pionteck2021}, and type-B aortic dissection \cite{Chen2018,Kan2021}. By taking personalized information as input, modeling also supports the design of patient-specific medical implants \cite{McCloy2001}. 

For in silicio modeling of personalized stent intervention in CoA, a protocol for virtual geometry deformation \cite{Neugebauer2016} and a validated numerical method to accurately predict the blood flow in the aorta are required. Regardless of the erythrocytes, leukocytes, and platelets in blood, the flow in the aorta is normally modelled as Newtonian fluid \cite{pedley1980} considering the relatively large Reynolds number $Re$, which is proportional to the flow velocity and aorta diameter and inversely proportional to the blood viscosity. Computational fluid dynamics (CFD) plays an important role in resolving hemodynamics \cite{Taha2005}. Due to the personalized and complex 3D geometry and jet flows induced by heart contraction and local narrowing, laminar flow, turbulent flow and transition between them may coexist spatiotemporally \cite{Stein1976, Ku1997}. Thus, to accurately resolve such aortic flow, both turbulence and complex geometry should be considered in CFD simulations. Three approaches, including Reynolds-averaged Navier–Stokes equations (RANS), large eddy simulation (LES), and direct numerical simulation (DNS), are typically used for turbulence modeling. From RANS to DNS, both the accuracy and computational demand increases due to more and more degrees of freedoms that need to be resolved.

So far, mainly RANS and LES were used to study aortic flow in literature \cite{Caballero2013}. With a transitional model, RANS was adopted to resolve flows in patient-specific thoracic aortic aneurysm by Tan et al. \cite{Tan2009} and aortic dissection by Cheng et al. \cite{Cheng2013,Cheng2015}. Simulations were carried out using ANSYS CFX, a commercial finite volume-based solver. Kouseral et al. studied flow stability in a normal aorta using the same numerical method, and compared their numerical results with experimental data from in vivo magnetic resonance imaging (MRI) \cite{Kousera2012}. They concluded that the RANS based shear stress transport transitional model was capable of capturing the correct flow state when low inflow turbulence intensity (1.0\%) was specified. Miyazaki et al. validated three CFD models for aortic flows in the aorta of a healthy adult and a child with double aortic arch \cite{Miyazaki2017}. Laminar, LES and the renormalization group (RNG) \textit{k}-$\varepsilon$ model were considered and compared. Simulations were performed using another finite volume-based solver, ANSYS Fluent. Their results show that the RNG \textit{k}-$\varepsilon$ model has the highest correlation with data from 4D flow MRI. Recently, Manchester et al. used LES to study the blood flow in patient-specific aorta with aortic valve stenosis \cite{Manchester2021}. Here, the finite volume based open-source library OpenFOAM, was used. After investigating the fluctuating kinetic energy, wall shear stress (WSS) and energy loss, they concluded that turbulence played an important role in aortic hydrodynamics. 

It should be noted that severe turbulence will be encountered in CoA due to a more complex geometry and larger $Re$, which might lead to higher requirements on the CFD method. The aforementioned conventional CFD methods are based on discretizations of macroscopic governing equations, such as the Navier-Stokes (NS) equations. Alternatively, the lattice Boltzmann method (LBM) is based on the mesoscopic Boltzmann equation and has multiple advantages, including simply handling of complex geometric shapes, ease of programming, and suitability for parallelization \cite{kruger2017lattice, he2009lattice, chen1998lattice}. Therefore, the LBM is increasingly used for the simulation of turbulent flow \cite{Chen2003} and biological fluid flows \cite{Wang2014}. Hennt et al. simulated the unsteady blood flow in a patient-specific geometry with a moderate thoracic aortic coarctation, and demonstrated that the LBM based DNS was capable of resolving such complex flow \cite{Henn2013}. Recently, Mirzaee et al. studied aortic flows for 12 patients with CoA using the LBM based LES, particularly with the Smagorinsky turbulence model \cite{Mirzaee2017}. A reasonable agreement for pressure drop between the numerical results and the catheter measurements was achieved. Nevertheless, to guide in silico stent intervention for CoA, a comprehensive validation for the LBM based LES for complex flow is still missing.

Since the 1970s and 1980s, MRI has become an important clinical and scientific tool that is widely used for diagnosis, monitoring of treatment procedures, and for biomedical research \cite{Markl2012}. Compared with X-ray and computed tomography, one of the advantages of MRI is the use of non-ionizing radiation \cite{McRobbie2007, Landheer2020}. In addition to obtaining anatomical information, MRI can also be used for quantitative flow measurements using phase-contrast imaging \cite{Moran1982, DonnellM1985,Soulat2020} including measurement of aortic blood flow \cite{Miyazaki2017,Saitta2019}.

In this study, the LBM based LES and DNS were adopted to resolve blood flow in
the human aorta. Geometries for a patient-specific aorta with CoA, before and
after stent intervention, were considered and physical phantoms were created
using 3D printing for use in MRI flow experiments. Flow measurements obtained
with MRI scans were then used as boundary conditions for simulations. Obtained
numerical results using LES and DNS were then compared with experimental 
4D-Flow data. To further validate the LBM based LES, we also compared with
in vivo data. We demonstrated that LES is capable of accurately simulating 
complex aortic flow and further applied it for in silico stent implantation.
Details of the methodology are given in Section 2. Numerical results and
experimental measurements for aortic flows are presented and comapred in Section
3. The application for stent selection is provided in Section 4.
Discussion and conclusions can be found in Section 5.

\section{Methodology}

\subsection{MRI Experiments}

The anatomical structure of the aorta and the flows therein were acquired by
MRI, which provided realistic geometries and boundary conditions for CFD
simulations. Comparison between CFD and phase-contrast flow MRI for flows in
3D-printed phantoms and in vivo aorta were performed respectively. 

\begin{figure}
\centering
\center{\includegraphics[trim={0 0 0 0}, clip, width=0.9\linewidth]{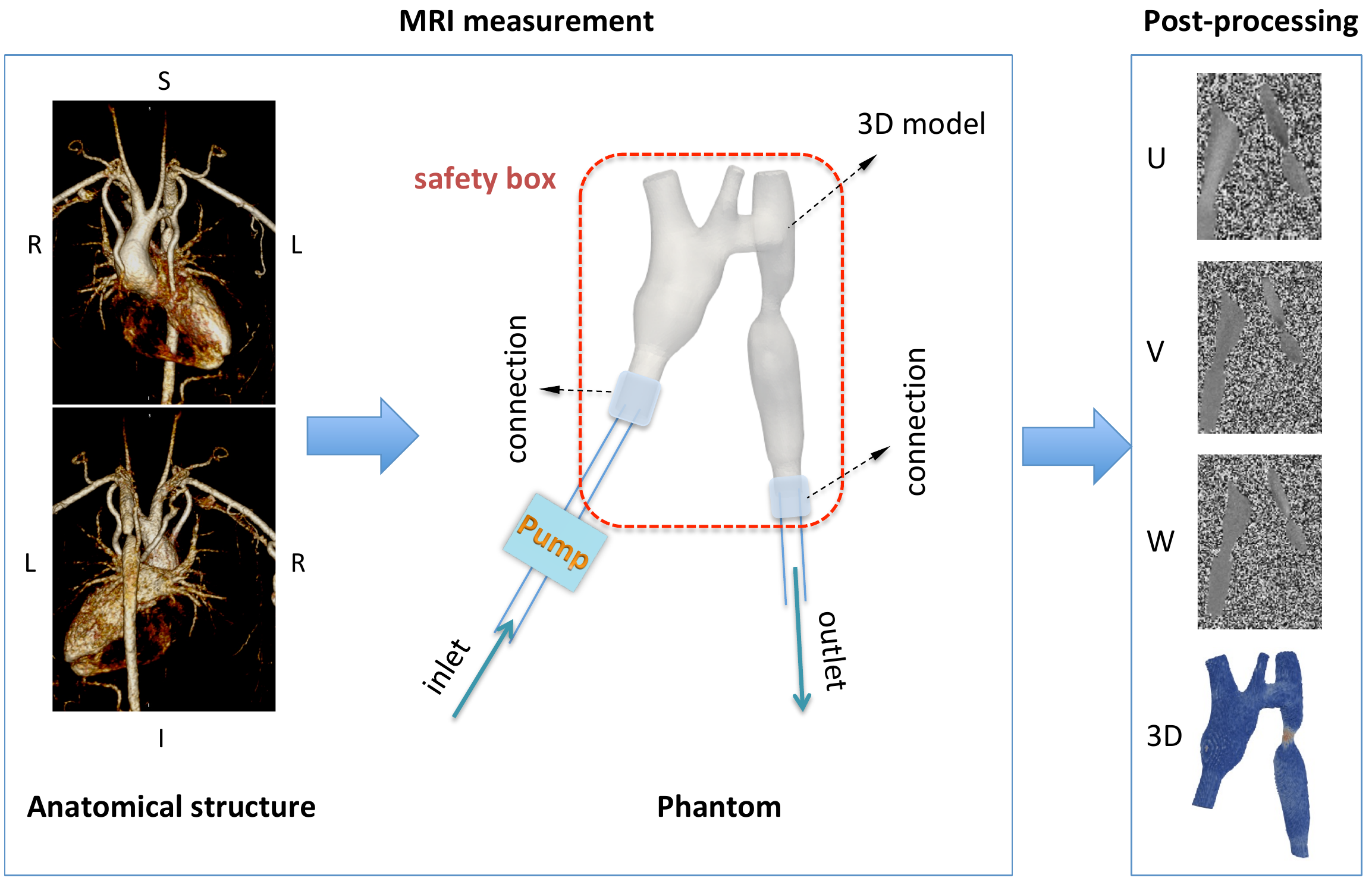}} 
\caption{Schematic diagram of the phantom measurement. Left: anatomical 3D
structures of a heart and great vessels (with CoA) were segmented and
reconstructed based on MRI images to obtain the geometry of the aorta.
Middle: 4D-Flow MRI of 3D printed phantoms
connected to a pump were performd for pre-interventional  and
post-interventional geometries. Right: 4D-Flow MRI yields three
components (U, V, W) of the instantaneous flow velocity at each voxel.} 
\label{fig:FlowPhatom}
\end{figure}

For the phantoms, the 3D anatomies of the heart and aorta of a 14-year-old
patient with CoA, before and after stent intervention, were reconstructed from
images obtained by a Magnetom Skyra 3T (Siemens Healthineers, Erlangen,
Germany). The stent (diameter = 12 mm) used in this patient was a covered 
Cheatham-platinum (CP) stent made of platinum-iridium (NuMed, Orlando, USA).
The sequence parameters
are listed in Table 1. Using ITK-SNAP \cite{YUSHKEVICH20061116}, the geometry
that starts from the aortic root and ends above the diaphragm was segmented
based on the grey values and exported as STL file. The main branches, such as
the right subclavian, the left subclavian, the right carotid artery and the
left carotid artery, were included. To have an uniform surface mesh, the
generated geometries were then remeshed using Autodesk Meshmixer
\cite{SchmidtR2010}. The schematic diagram of the experiment is shown in Fig. 
\ref{fig:FlowPhatom}. Two aortic models, including the pre-interventional and
post-interventional geometries, were printed using the Stratasys’ high-end 3D
laser printer Connex 3 using biocompatible MED610 as material. The phantoms
were connected to a pump. Forced water flows therein were then measured using
4D phase-contrast Flow MRI \cite{Markl2012, Dyverfeldt2015, Potters2014}.
The sequence parameters can be found in Table 1. Every case was measured three
times with about 30 mins per measurement. The averaged flow fields were then
used for comparison.
                    
For the in vivo validation, the aortic blood flow of a 3-year-old patient,
was obtained using a 2D flow sequence instead of a 4D flow sequence, to reduce the
duration of measurement. In the 2D measurement, the through-plane velocity of
the flow  was measured in two planes located in the ascending aorta and
descending aorta respectively. All in vivo measurements were made with the use
of ECG triggering and respiratory gating. More details can be found in Table 2.
For further CFD simulation, the aortic geometry was segmented and reconstructed
using the same procedure as mentioned above.

% --------------- [table ↓] --------------- %
\begin{table*}
\small
\caption{MRI sequences used for the flow-phantom study:
In vivo MRI scans were performed pre and post intervention
to obtain the anatomical structure. These geometries were
used for 3D-printing a phantom then used
for 4D-flow MRI measurements.}

\begin{center}
\begin{tabular}{c|cccc}
\hline\hline
          & Geometry (pre) & Geometry (post) & 4D flow MRI (pre) & 4D flow MRI (post) \\
\hline 
Sequence type &{3D FLASH (TWIST)} &{3D T1 weighted FLASH} &3D Cartesian FLASH & 3D Cartesian FLASH \\
Acceleration & 3 $\times$ 2 & 2 $\times$ 2 & - & - \\
Matrix size &352 $\times$ 246 &448$\times$252  &384 $\times$504  &416$\times$364\\
Number of slices  &80 &88 &144 &144\\
Slice thickness (mm)  &1.30 &1.20 &0.77 &0.77\\
Pixel size (mm$^2$) &1.02$\times$1.02 &0.89$\times$0.89 &0.77$\times$0.77 &0.77$\times$0.77\\
Repetition time (ms)  &2.75 &3.70 &36.40 &70.40\\
Echo time (ms)  &1.00 &1.31 &4.61 &7.46 \\
Flip angle ($^\circ$)  &20 &25 &7 &7\\
Velocity encoding (cm/s)  &- &- &50 &40\\

\hline\hline
\end{tabular} 
\label{tab:sequences}
\end{center}
\end{table*}

% --------------- [table ↓] --------------- %
\begin{table*}
\small
\caption{MRI sequences for geometry and 2D flow 
for in vivo validation.}
\begin{center}
\begin{tabular}{c|cc}
\hline\hline
          & Geometry & 2D flow MRI\\
\hline
Sequence type &{3D T1 weighted FLASH} &{2D T1 weighted FLASH}\\
Acceleration & 3$\times$2 & 2 $\times$ 2  \\
Matrix size &320$\times$260  & 192$\times$119\\
Number of slices  &88 &30 \\
Slice thickness (mm)  &1.00 &5.00\\
Pixel size (mm$^2$) & 1.00$\times$1.00 & 1.56$\times$1.56\\
Repetition time (ms)  & 312.01&39.44\\
Echo time (ms)  &1.64 &2.67\\
Flip angle ($^\circ$)  &20 &20\\
Velocity encoding (cm/s)  &- &200\\

\hline\hline
\end{tabular} 
\label{tab:sequences}
\end{center}
\end{table*}

\subsection{Numerical Modeling}

The CFD simulations in this study were performed using the LBM, which is based
on the kinetic theory, particularly the Boltzmann equation which describes the
movements of fluid particles \cite{kruger2017lattice, he2009lattice,
chen1998lattice,He1997}. For simulations, the space and time are discretized into
finite nodes and time steps. Starting from an initial state, the configuration
of the fluid particles at each time step evolves in two sub-steps, streaming
and collision. During streaming, fluid particles at a node move to the
neighbouring nodes along specified discrete directions as defined by
the lattice. The streamed particles at a node collide with each other and
change their velocity distribution functions \cite{Roberto1992}. For 3D flows,
the most popular lattice is the D3Q19, which is used in this work. 

Different operators, such as the single-relaxation time BGK \cite{Qian1992} and
the multi-relaxation-time (MRT) operators \cite{DHumieres2002}, can be used to
approximate the particle collision. We chose the MRT operator due to its better
numerical stability. The governing equation for the LBM with MRT operator reads
\begin{equation}
f_{i}\left(\mathbf{x}+\mathbf{e}_{i} \Delta t, t+\Delta t\right)-f_{i}(\mathbf{x}, t)=\Lambda_{ij} ( f_j^{eq}(\mathbf{x}, t)-f_j(\mathbf{x}, t)), 
\label{eq:lbm}
\end{equation}
in which $f_{i}$ is the particle velocity distribution function along the $i$th
direction; $\mathbf{x}$ and $t$ are the spatial coordinate and time
respectively; $\Delta t$ is time step; $\mathbf{e}_{i}$ is the discrete
velocity of the lattice along the $i$th direction. The right-hand side of Eq.
(\ref{eq:lbm}) represents the collision process in momentum space.
$\Lambda_{ij}=\mathbf{M}^{-1}\mathbf{S}\mathbf{M}$; $\mathbf{M}$ is a given
transformation matrix for the lattice; $\mathbf{S}$ is a diagonal matrix.
Macroscopic parameters, such as the fluid density, pressure and velocity, are
moments of $f_i$. 

The left-hand side and right-hand side of Eq. (\ref{eq:lbm}) represent the
streaming and the collision processes respectively. The simplicity of this
equation implies that the LBM is readily parallelizable as the non-local
streaming is linear while the non-linear collision is local
\cite{kruger2017lattice, he2009lattice, chen1998lattice}. Thus, the LBM is
increasingly used for turbulence modeling, especially DNS with high performance
modern computers. Additionally, due to its particle feature, even with a simple
Cartesian grid the LBM can resolve flow with complex geometry, such as the
patient-specific aortas considered in this study.

The LBM based DNS and LES were investigated in this study, based on the
open-source library Palabos \cite{LATT2021334}. DNS resolves the flow at all
scales without empirical model, thus is regarded as a kind of numerical
experiment. As mentioned before, the calculation cost of DNS is very high
especially for flows with large Reynolds number \cite{Moin1998,Wang2014}.
Alternatively, LES explicitly solves large eddy current and implicitly
calculates small eddies by using a sub-grid scale (SGS) model, thus balancing
accuracy and computational cost. The Smagorinsky SGS model \cite{Joseph1963}
was incorporated into the LBM in this study. 

For all simulations, the inlet velocity with Poiseuille profile was specified
at the ascending aorta. The flow rates were given based on MRI measurements and
can be found in the following sections. Outlet boundary condition with a
reference pressure was applied to the descending aorta. The curved aortic wall
was assumed to be no-slip and treated with an extrapolation scheme
\cite{Guo2002}.

\section{Validation and Comparison}

\subsection{Phantom Experiments}

The two phantoms filled with water were used in the MRI
experiments and compared to CFD simulations of the same geometries.
The main branch blood vessels were closed, to
reduce their influence. Inlet and outlet of the geometries were extended
artificially for the connection of the water pipe. Water is incompressible
and Newtonian. Its density and kinematic viscosity are 1.0 kg/m$^{3}$ and 1.0
$\times 10^{-6} \text{m}^2\text{/s}$, respectively. The averaged velocity at
the inlet is 0.1 m/s.  Different spatial resolutions were considered for every
single case. After mesh independence tests, 12.30 million (DNS) and 3.45
million (LES) lattice nodes were selected for the pre-interventional geometry
and 5.12 million (DNS) and 1.25 million (LES) lattice nodes for the
post-interventional one. 

\begin{figure}
    \centering
    \center{\includegraphics[width=0.9\linewidth]  {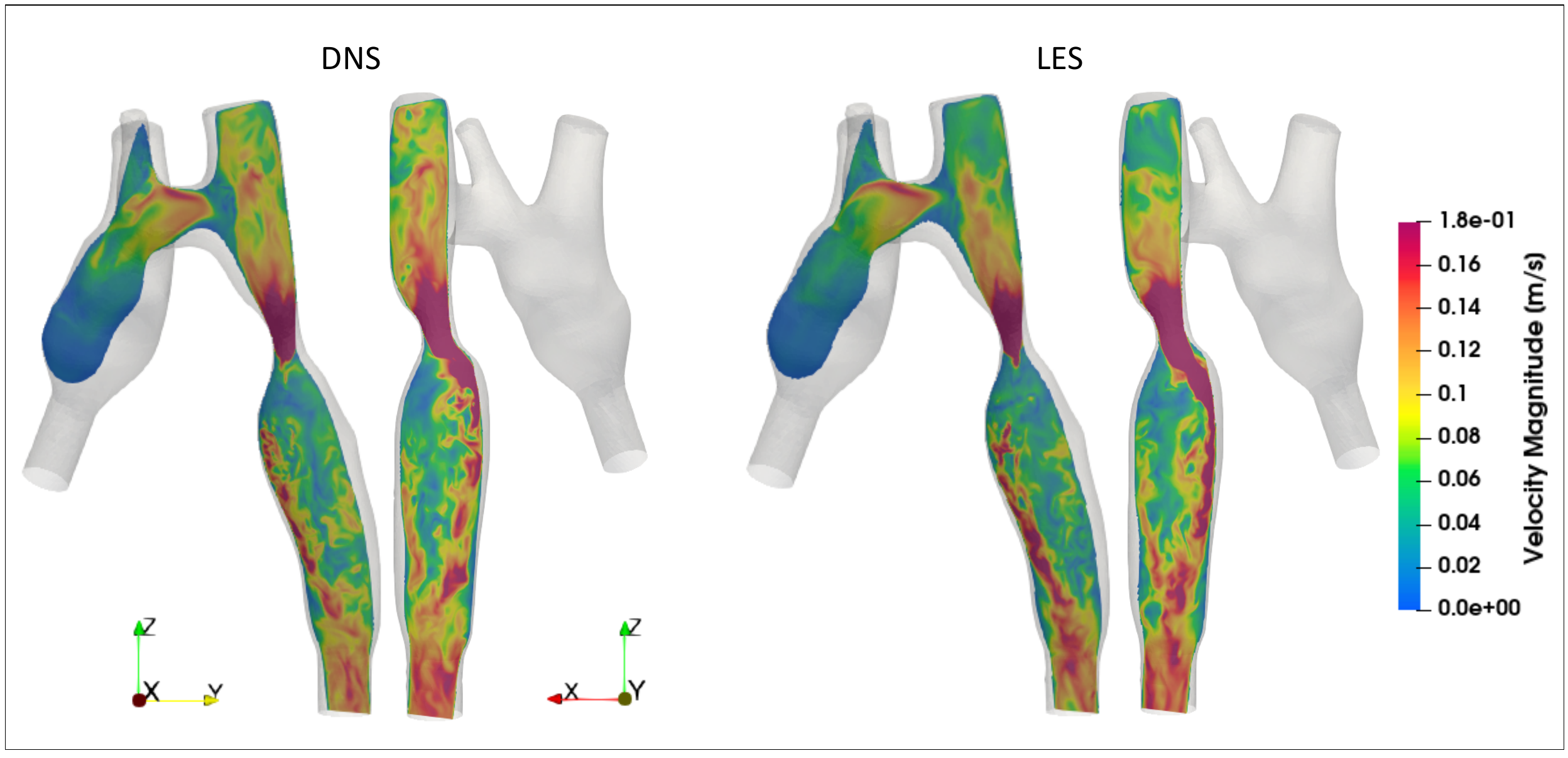}} 
    \caption{Instantaneous velocity contours on a sagittal plane and a coronal plane
	in the pre-interventional geometry of an aorta of a patient with CoA
	simulated by CFD. Left: DNS; right: LES.}
 \label{fig:phantom-instantFields.pdf}
\end{figure}

Due to the complex geometry, \textit{e.g.} multiple plane curvatures and
branches, blood flow in the patient-specific aorta is unsteady and complicated.
Instantaneous velocity contours on a sagittal plane and a coronal plane in the
pre-interventional geometry are given in Fig.
\ref{fig:phantom-instantFields.pdf}. Due to the relatively low temporal
resolution of 4D flow MRI, here only results from DNS and LES are presented. It
can be seen that the flow therein is turbulent. Because of the local narrowing
in the stenosis, flow is accelerated in the pre-interventional geometry and the
local Reynolds number on the stenosis plane is more than 2500. Jet flow, which
leads to high blood pressure and high wall shear stress (WSS), is observed. DNS
provides more flow details due to higher spatial resolution. 

\begin{figure}
    \centering
    \center{\includegraphics[width=0.9\linewidth]  {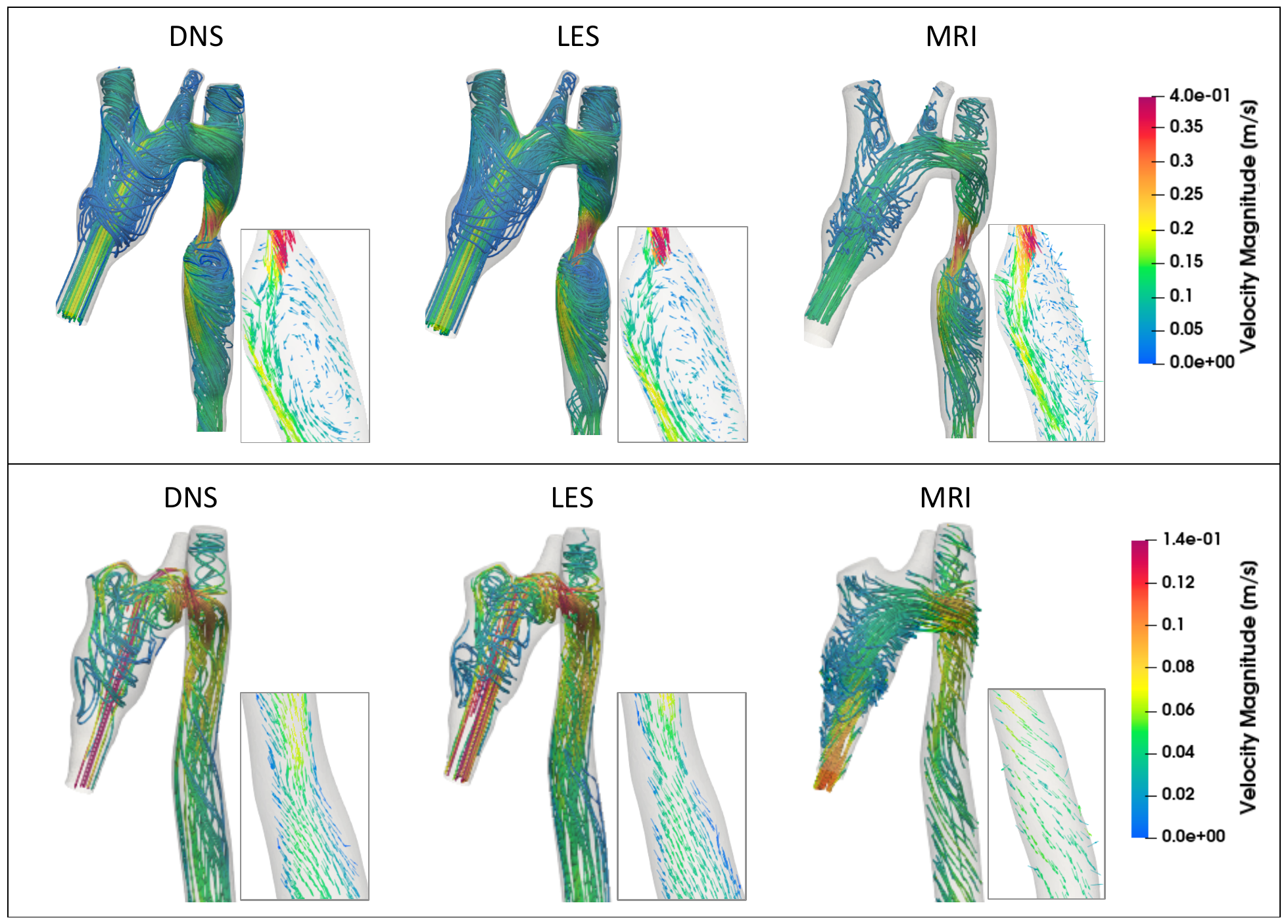}} 
    \caption{Streamlines of the flow computed with CFD simulations of the aorta
	compared to MRI measurements of the flow in the 3D-printed phantoms. 
	The time-averaged flow is shown. The insets show the velocity vectors on
	the sagittal planes. Top: pre-interventional
	geometry; bottom: post-interventional geometry.}
	\label{fig:phantom-streamlines.pdf}
\end{figure}

As the flow is unsteady, temporal averaging was performed for both CFD and MRI results
and the following comparison is based on the time-averaged flow fields.
The main flow features can be found in Fig. \ref{fig:phantom-streamlines.pdf}.
The visualization of 3D streamlines and velocity vectors on the sagittal plane
(insets) shows the complexity of the flow within the patient’s aorta,
especially where the stenosis occurs. Again, jet flow and recirculation are
observed in the pre-interventional geometry in the streamlines and highlighted
in the zoomed-in insets. Helical streamlines can also be found in all cases.
For the MRI results, some streamlines start from the vessel wall as no-slip
boundary condition is not guaranteed in MRI data. Nevertheless, all methods,
including DNS, LES and MRI, resolved the main flow features. Moreover, as the
aorta is deformed and flattened after stent implantation, the flow resistance
in the post-interventional geometry is reduced. The pressure drop is reduced
from 790 Pa (DNS) and 778 Pa (LES) to 9 Pa (DNS) and 8 Pa (LES), respectively.
Those results indicate that stent implantation restored the aortic flow
effectively.

\begin{figure}
    \center{\includegraphics[trim={0 0 0 0}, clip, width=0.9\linewidth]
    {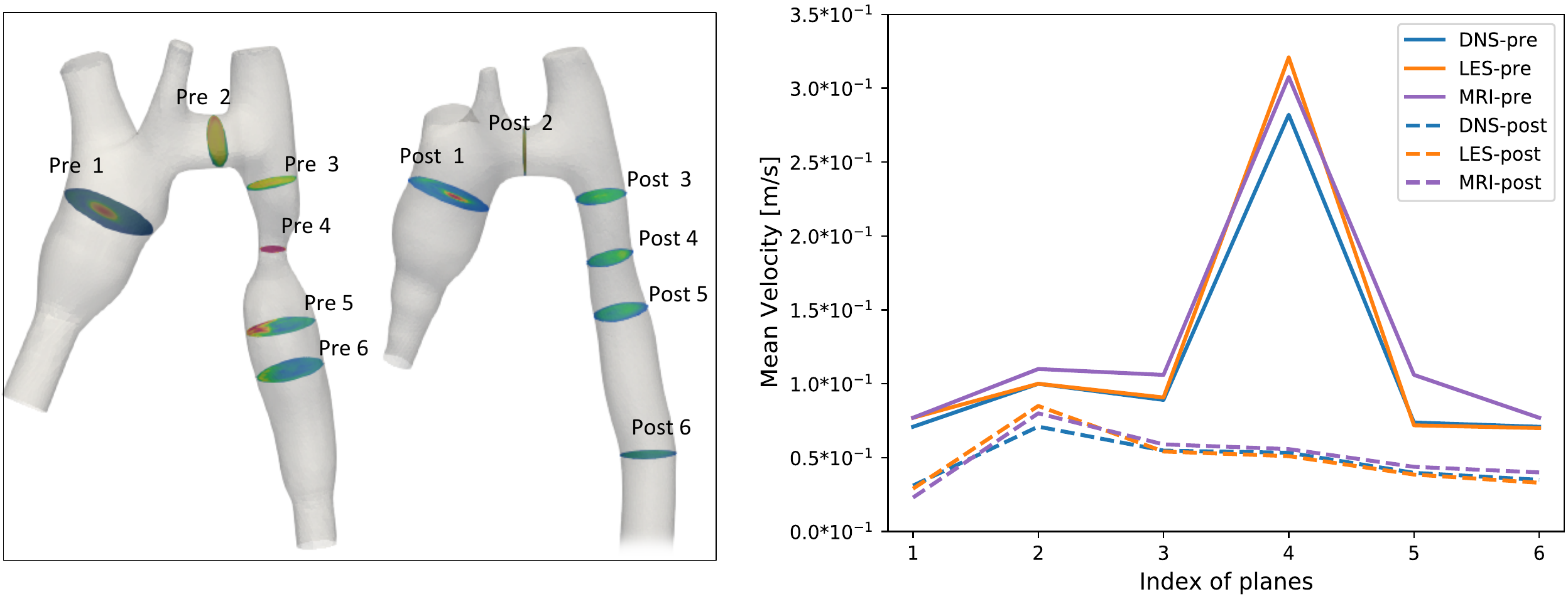}} 
    \caption{Comparison of mean velocity magnitude on specified planes
	for the pre-interventional and post-interventional aorta geometry.
	Left: six specified cross planes, including ascending, arch, pre-stenosis,
	on-stenosis, post-stenosis and descending of aorta. Right: The
	time-averaged mean velocities from 4D flow MRI and the CFD simulations.} \label{fig:MeanVelocityofSixPlanes}
\end{figure}

Figure \ref{fig:MeanVelocityofSixPlanes} presents quantitative comparison of
mean velocity magnitude on six specified cross-sectional planes. These six
planes, as shown in the left panel of Fig. \ref{fig:MeanVelocityofSixPlanes},
represent the ascending, arch, pre-stenosis, on-stenosis, post-stenosis and
descending of the aorta respectively. The mean velocity magnitude was
calculated according to $ \sum_{i=1}^{N}\sqrt{U_{i}^{2}+V_{i}^{2}+W_{i}^{2}}$ 
$/N $, with $N$  the number of points on a cross plane. It can be seen from the
right panel of Fig. \ref{fig:MeanVelocityofSixPlanes} that the MRI results are
a little larger than the numerical ones on planes Pre 2, Pre 3 and Pre 5. Using
MRI results as reference, the relative deviations for LES and DNS are $4-28\%$
and $7-27\%$ respectively. Largest deviation can be found on Pre 5, mainly due
to the difficulty for the MRI measurement induced by the recirculation after
the stenosis. On Pre 4, MRI data is between the DNS ($-4\%$) and LES ($7\%$)
ones. For the post-interventional geometry, CFD results agree with the
experimental data well on all planes, with relative derivations less than
10$\%$.

\begin{figure}
    \centering
    \center{\includegraphics[width=0.9\linewidth]  {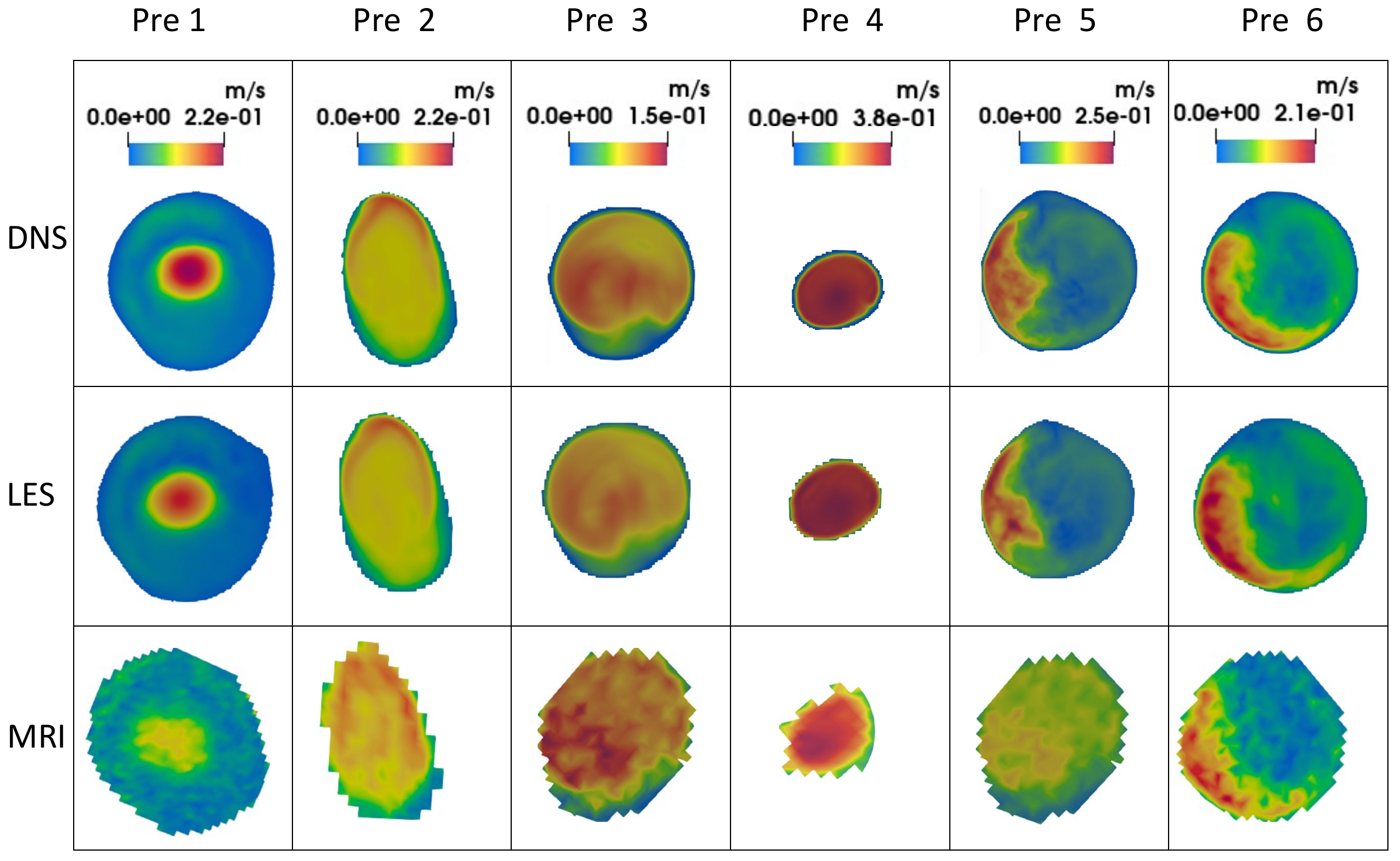}} 
    \caption{Flows in pre-interventional aorta. Velocity magnitude distributions were compared among CFD simulations and 4D Flow MRI on six planes as shown in Fig. \ref{fig:MeanVelocityofSixPlanes}. Time-averaged results.}
 \label{fig:pre-planes-VelocityMagnitude.pdf}
\end{figure}

\begin{figure}
    \centering
    \center{\includegraphics[width=0.9\linewidth]  {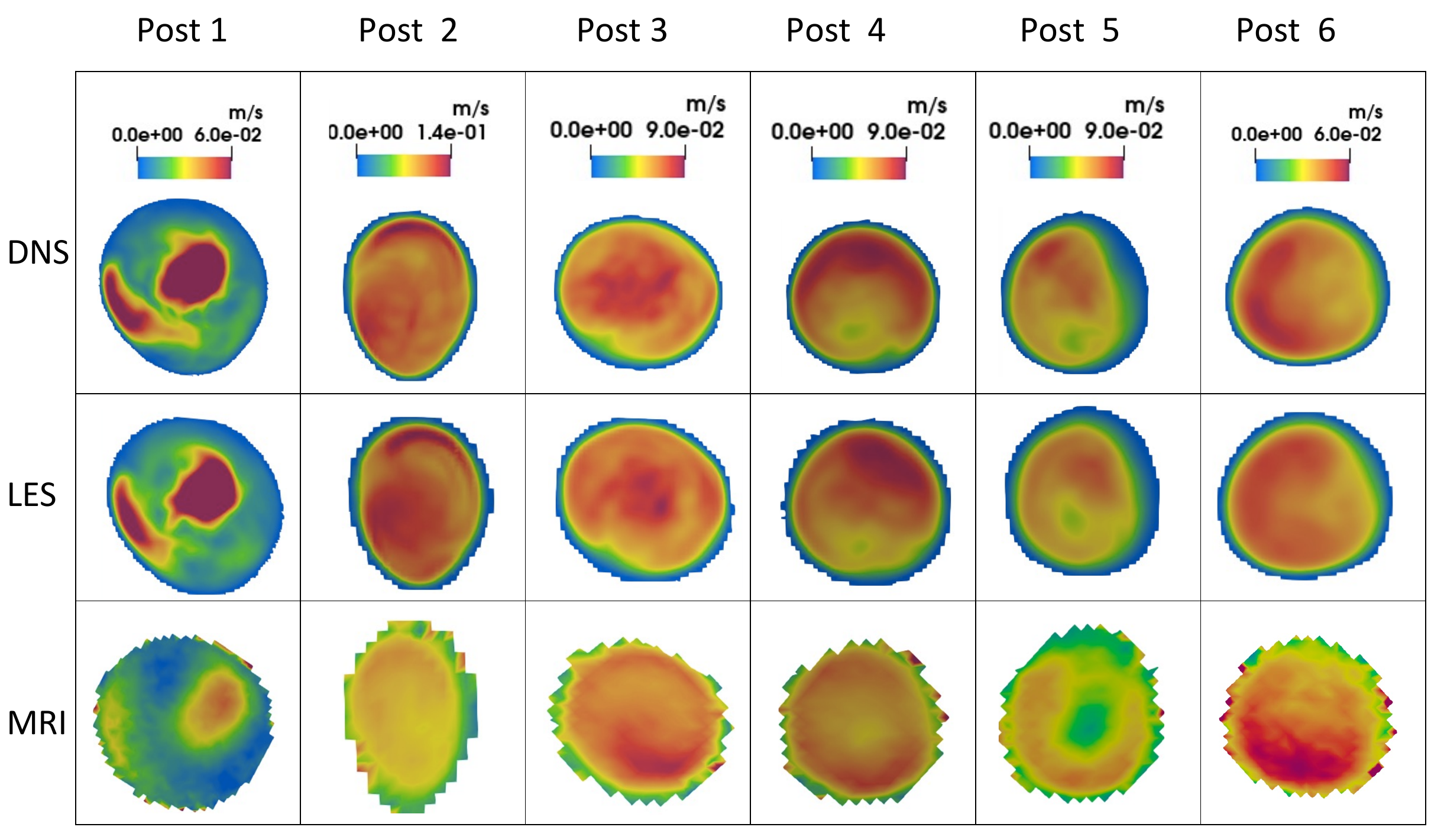}}
    \caption{Flows in post-interventional aorta. Velocity magnitude distributions were compared among CFD simulations and 4D Flow MRI on six planes as shown in Fig. \ref{fig:MeanVelocityofSixPlanes}. Time-averaged results.}
 \label{fig:post-planes-VelocityMagnitude.pdf}
\end{figure}

Velocity contours on those specified planes, in pre-interventional and
post-interventional geometries, are given in Figs.
\ref{fig:pre-planes-VelocityMagnitude.pdf} and
\ref{fig:post-planes-VelocityMagnitude.pdf} respectively. Generally, flow
patterns are qualitatively well-matched among LES, DNS and MRI for both
geometries. Both LES and DNS provide more details and smoother fields due to
higher spatial resolution compared with MRI. Discrepancy between the CFD
results and MRI one is observed on Pre 4 and 5. We believe this is because of
the complexity induced by the jet flow and recirculation after the stenosis,
which is visualized in Figs. \ref{fig:phantom-instantFields.pdf} and
\ref{fig:phantom-streamlines.pdf}. Nevertheless, the results from LES agree
very well with those from DNS on all specified planes.

\begin{figure}
    \centering
    \center{\includegraphics[width=0.9\linewidth] {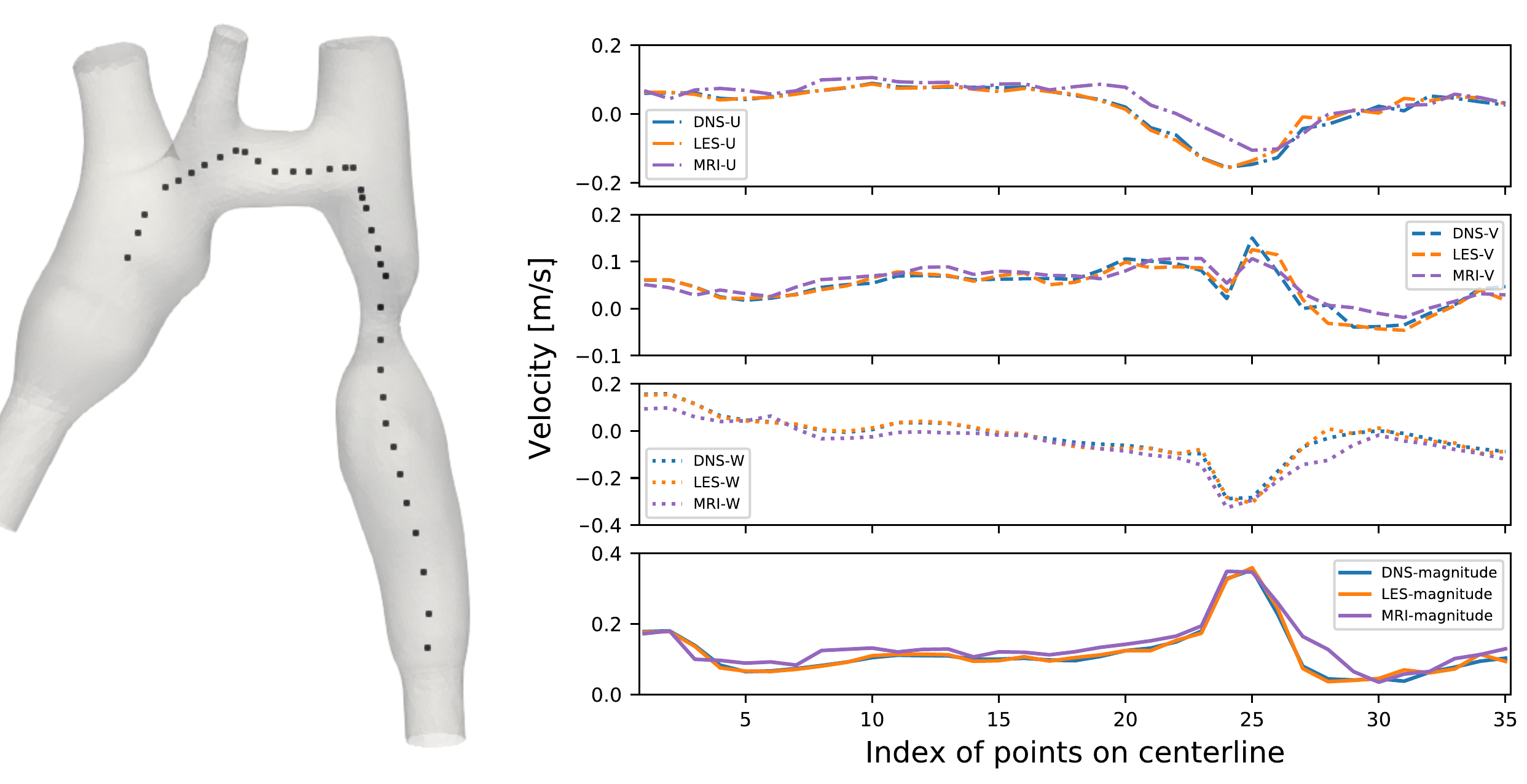}}
    \caption{Comparison of flow velocity along the centerline of the pre-interventional aorta. Left: locations of the points starting from the ascending aorta. Right: from top to bottom, velocity components and magnitude. Time-averaged results.}
 \label{fig:Pre-centerline-U_V_W.pdf}
\end{figure}

Velocities along the centerlines of the pre-interventional and
post-interventional geometries were also compared. The centerline was extracted
using the VMTK extension for 3D Slicer \cite{PINTER201919}. Since velocity in
the pre-interventional geometry changes more acutely because of the stenosis,
here only the results for pre-interventional geometry are presented. Discrete
points along the centerline, starting from the ascending aorta after the
artificial extension, are considered and can be found in the left panel of Fig.
\ref{fig:Pre-centerline-U_V_W.pdf}. Velocity components and magnitude are given
in the right panel. It can be seen that velocity components change sharply in
the stenosis region which leads to a peak in the profiles of velocity
magnitudes. Results from LES, DNS, and MRI agree with each other in most region
of the aorta, except the coarctation. Similarly to Fig.
\ref{fig:pre-planes-VelocityMagnitude.pdf}, discrepancy mainly happens in the
MRI result around the stenosis, while good agreement between DNS and LES is
always achieved.

Based on the above comparisons, it can be concluded that CFD (LES and DNS)
results agree with the data from 4D-Flow MRI. More information, such as
pressure drop and WSS can be easily generated from the CFD results. Since LES
needs less computational resource but provides acceptable accuracy, we conclude
that LES is capable in resolving aortic flow and adopt it for the following
research.

\subsection{In vivo Validation}

\begin{figure}
    \centering
    \center{\includegraphics[trim={0 0 0 0}, clip, width=0.9\linewidth] {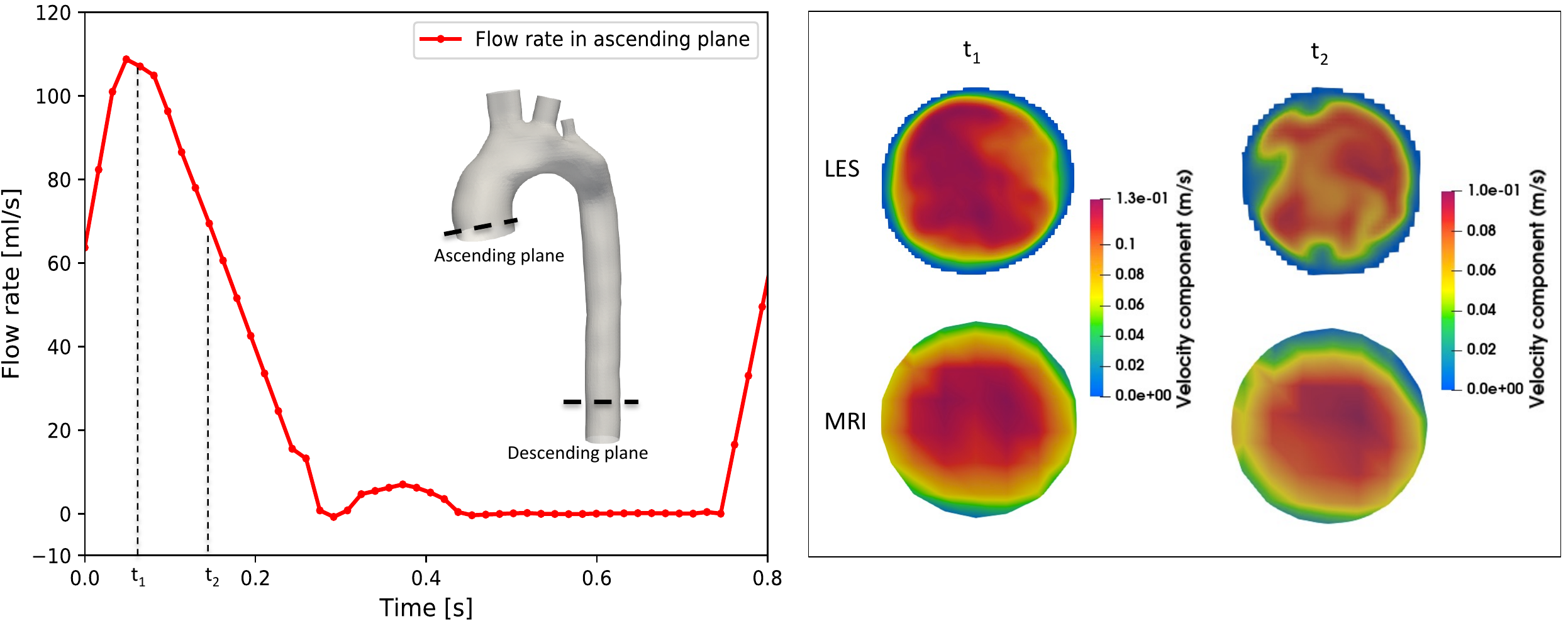}} 
    \caption{Comparison of in vivo aortic flow between LES and 2D flow MRI.
	Left: the experimentally recorded flow rate crossing the ascending
	plane; locations of the specified ascending plane and descending plane
	are given in the inset. Right: he through-plane component of velocity distributions on the descending plane at two different time instants during the cardiac cycle. }
	\label{fig:invivo.pdf}
\end{figure}

To further validate the numerical modeling, we also used in vivo measurements
in addition to the phantom study. The patient-specific aortic flow was compared
between LES and flow MRI. Considering the long measurement duration needed by
4D-Flow MRI \cite{Diagnosis2014}, 2D flow MRI was used for the in vivo scans.
Velocities, thus flow rates, on specified planes perpendicular to the main flow
direction were measured with 2D phase-contrast flow MRI. The experimentally
obtained flow rate on a plane in the acceding aorta is shown in the left panel
of Fig. \ref{fig:invivo.pdf}. The two time instants, $t_1$ and $t_2$ during
systole, were considered and their flow rates were used as inlet boundary
condition for the LES simulations. The experimentally recorded velocity
distributions in the descending aorta at these two instants were used for
comparison. 

In numerical simulations, the aortic geometry was segmented based on the MRI
slices. We didn't consider the deformation of the aorta, or fluid-solid
interaction, and assumed that the wall was not moving. A parabolic velocity
profile based on the experimentally recorded flow rate was specified as inlet.
The opening in the descending aorta was defined as outlet with a reference
pressure. Since the vessel branches were open in this test, the difference of
flow rates between the inlet and outlet were assigned to the branches according
to their cross areas \cite{Mirzaee2017, Pirola2017}. 

The time-averaged flow fields are given in the right panel of Fig.
\ref{fig:invivo.pdf}. At $t_1$ (the instant with peak flow rate), the through-plane component of mean velocity from LES is 0.69 m/s, with relative deviation 8$\%$ in reference to the MRI result, 0.75 m/s. Similarly, at $t_2$, the through-plane component of mean velocities are 0.55 m/s (LES) and 0.49 m/s (MRI). Thus, good agreement between LES and MRI was achieved
also in vivo. Moreover, LES with proper boundary condition could provide more
flow details due to higher spatial resolution, as shown in both in vivo and
phantom tests.

\section{Application for Stent Selection}

\begin{figure} \centering
	\center{\includegraphics[width=0.9\linewidth]  {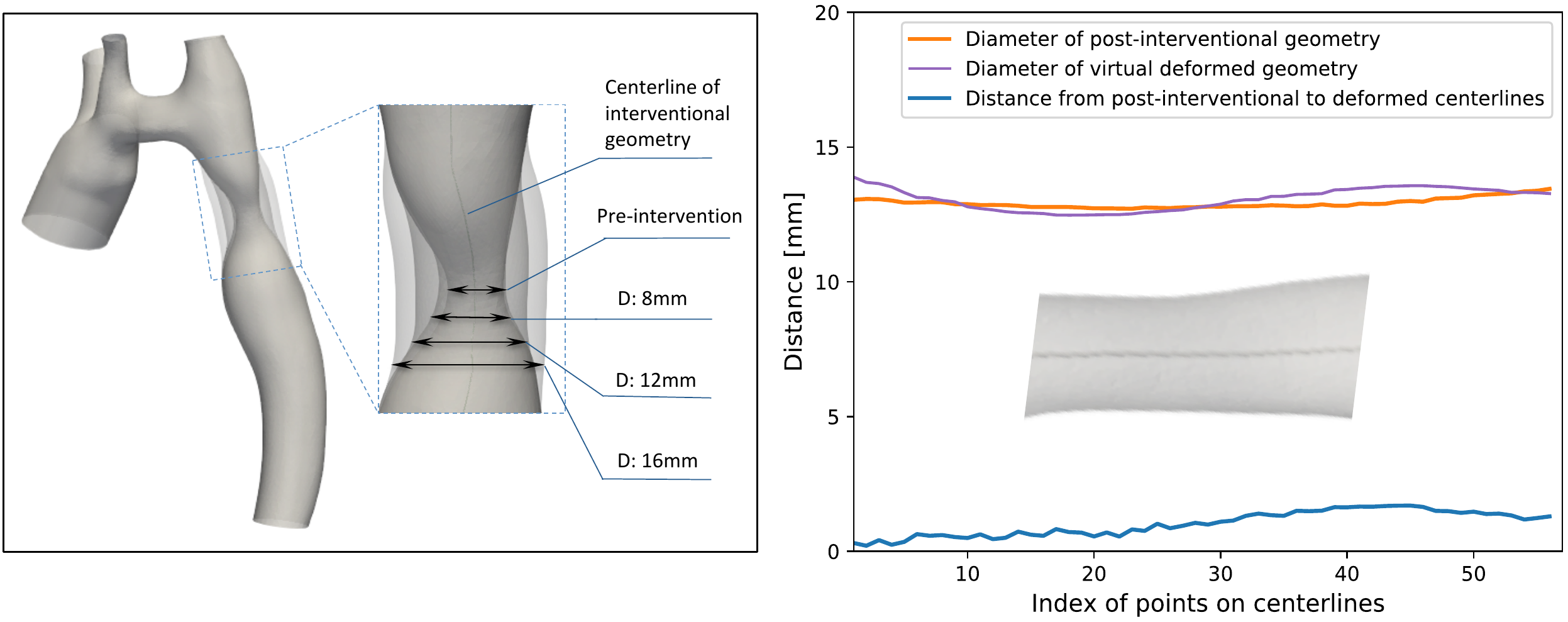}}
    \caption{Virtual stent implantation and validation of deformation. Left:
	pre-interventional geometry, virtually deformed geometries with stent
	diameter 8 mm, 12 mm, and 16 mm respectively. Right: validation of
	virtual stent implantation, with inset showing the realistic
	post-interventional geometry.} \label{fig:Deformed.pdf}
\end{figure}

The main motivation for silico stent implantation is to help clinicians to
evaluate the surgical plans based on predicted results and to be able to select
an optimal stent already before surgery. A fast virtual stenting approach
proposed by Neugebauer et al. \cite{Neugebauer2016} was implemented in this
work to generate virtually deformed geometries. Several parameters, such as the
aorta bending resistance, aorta stiffness, stent stiffness, stent position and
diameter were considered in this approach to represent the interaction between
aorta and stent. The deformed geometry deformation was then obtained based on
the deformed centerline and deformed surface vertices. The original
pre-interventional geometry and its deformed versions can be found in the left
panel of Fig. \ref{fig:Deformed.pdf}. The deformed geometries were exported as
STL files, which were remeshed and further imported into the CFD solver for
flow simulation. It should be noted that in this study the physician chose a
stent with diameter 12 mm independently and without input from the in slicio
model. We compared the virtually deformed geometry based on this stent with the
post-interventional one reconstructed from MRI images in the right panel of
Fig. \ref{fig:Deformed.pdf}. Centerlines for both geometries were obtained and
the distance ( $\leq 1.60$ mm) between two centerlines is presented. Similarly,
the spatial dependent diameters are also given, with deviation less than 0.65
mm. It can be concluded that the virtually deformed geometry agrees with the
clinically deformed one quantitatively. 

\begin{figure} \centering
    \center{\includegraphics[width=0.9\linewidth]  {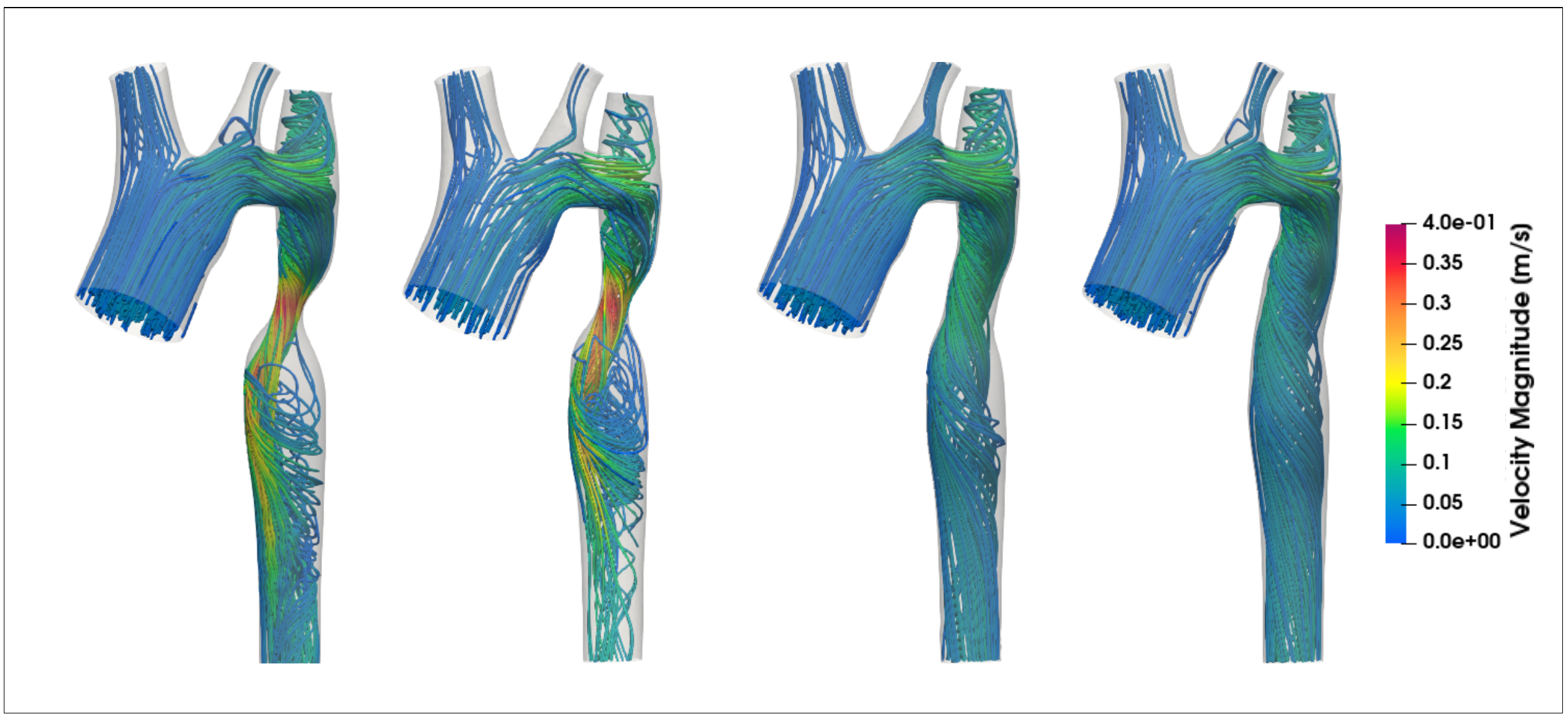}}
    \caption{Streamlines in the pre-interventional and virtually deformed
	aortas. Time-averaged results. From left to right: pre-interventional
	geometry, virtually deformed geometries with stent diameter 8 mm, 12
	mm, and 16 mm respectively.} \label{fig:deformedAorta-streamlines.pdf}
\end{figure}

\begin{figure}
    \centering
    \center{\includegraphics[width=0.9\linewidth]  {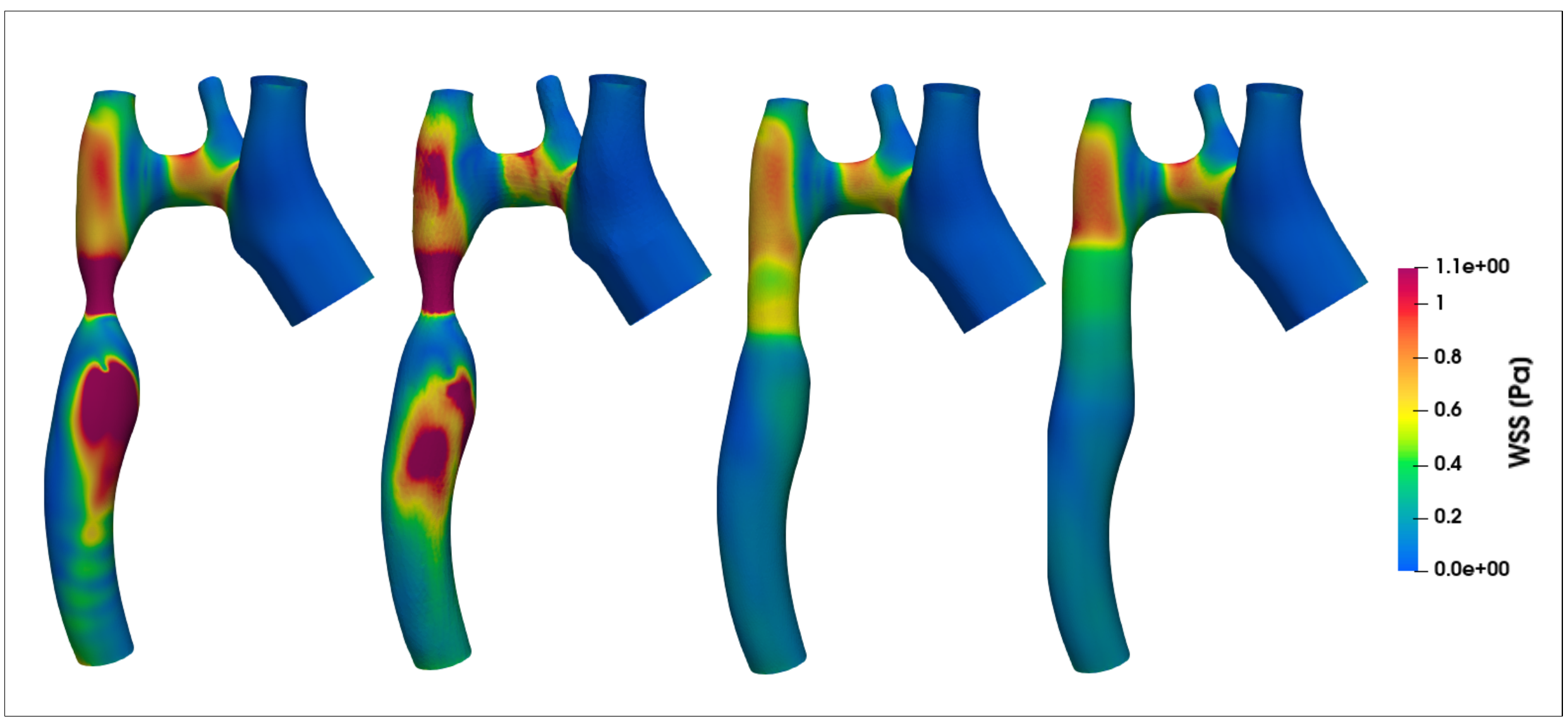}}
    \caption{WSS on the pre-interventional and virtually deformed walls.
	Time-averaged results. From left to right: pre-interventional geometry,
	virtually deformed geometries with stent diameter 8 mm, 12 mm, and 16
	mm respectively. } \label{fig:deformedAorta-WSS.pdf}
\end{figure}

The above validated LES was used to resolve the
flows in the pre-interventional geometry and the virtually deformed ones. We used
the same boundary conditions as mentioned in subsection 3.2, with flow rate
13.2 ml/s at the inlet. The numerically obtained results are presented in Figs.
\ref{fig:deformedAorta-streamlines.pdf}-\ref{fig:deformedAorta-WSS.pdf}. The
color coded streamlines in Fig. \ref{fig:deformedAorta-streamlines.pdf} show
that a stent with diameter 8 mm is inadequate to reduce the stenosis and jet
flow with a large local flow still can be observed in the narrowing region. For
stent diameter 12 mm or 16 mm, the jet flow disappears, with substantially
reduced maximum velocity compared to the pre-interventional geometry. The WSS
distributions are given in Fig. \ref{fig:deformedAorta-WSS.pdf}. WSS describes
the mechanical force generated by blood flow on the vessel wall, thus plays an
important role in chronic adaption and remodelling \cite{Humphrey2021}. It is
defined as $\tau_{y=0}= \mu \frac{\partial u }{\partial y}\mid  _{y=0} $, where
$\mu$ is the dynamic viscosity of the flow, $u$ is the flow velocity along the
wall and $y$ is the height above the wall.  As shown in Fig.
\ref{fig:deformedAorta-WSS.pdf}, high WSS is observed in the stenosis region in
the pre-interventional geometry and the deformed one with stent diameter 8 mm.
After the stenosis, high WSS is also found in a part of the descending aorta
due to the impact of high-speed jet flow (see Fig.
\ref{fig:deformedAorta-streamlines.pdf} for reference). A stent with diameter
16 mm enlarges the stenosis most and therefor leads to the smallest WSS in the
same region. However, as this stent is larger than the size of the aorta, it
also leads to a relatively large WSS before the stent, compared to the case
with stent diameter of 12 mm.

\begin{figure}
    \centering
    \center{\includegraphics[width=0.55\linewidth]  {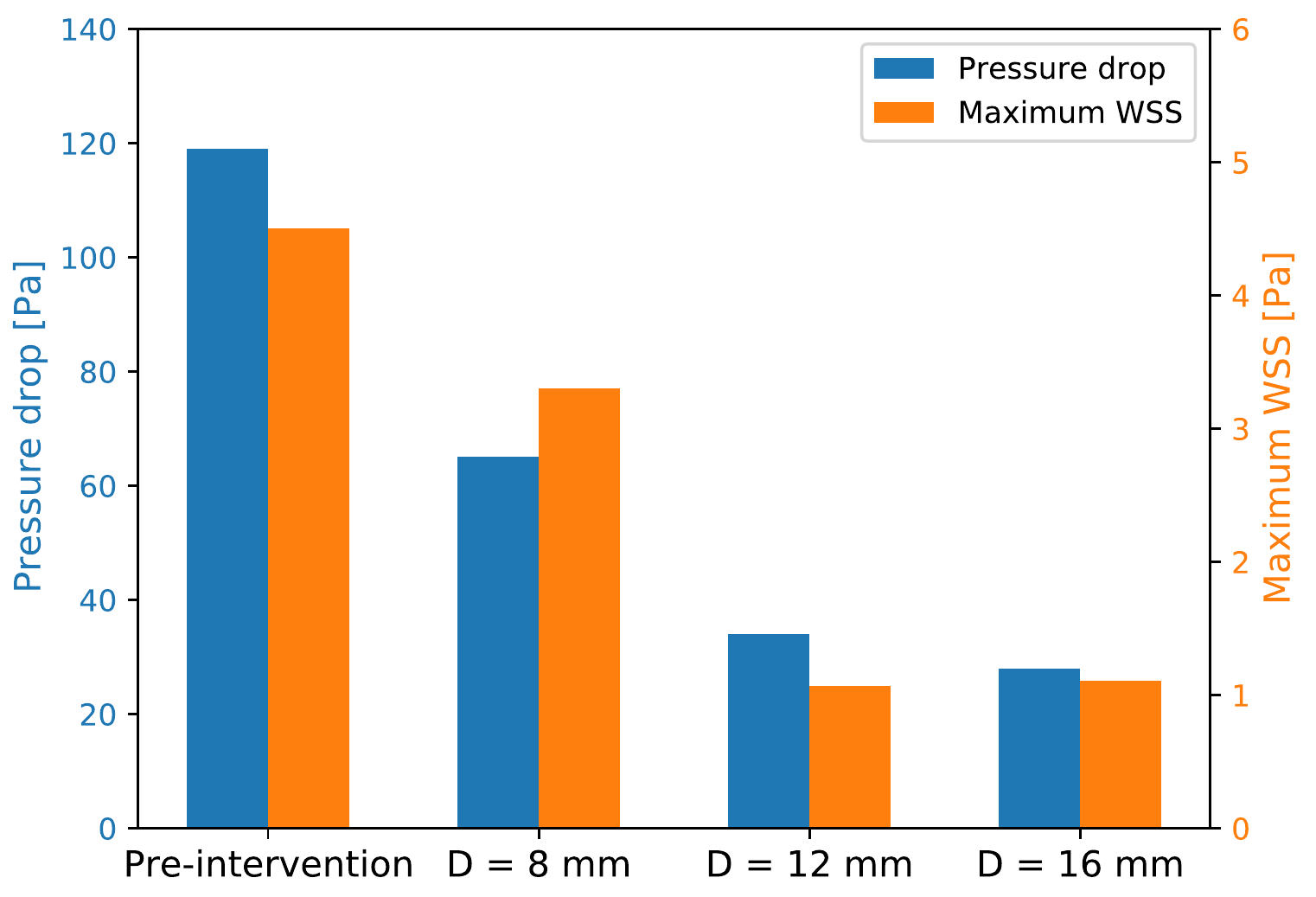}}
    \caption{Quantitative comparison of pressure drop and maximum WSS in the aortas before and after deformation.}
 \label{fig:deformedAorta-comparsion.pdf}
\end{figure}

To quantitatively compare the four cases with different stent diameter, the
pressure drop and maximum WSS are given in Fig.
\ref{fig:deformedAorta-comparsion.pdf}. It can be seen that the pressure drop
is 119 Pa in the pre-surgical geometry, and is reduced to 34 Pa and 28 Pa in
the geometry with stent diameter 12 mm and 16 mm respectively. For the maximum
WSS on the aortic wall, the geometry with stent diameter 12 mm provides the
smallest value, 1.07 Pa. It is understandable that a larger diameter stent
results in less flow resistance thus smaller pressure drop, assuming that the
aortic wall is always deformable. On the other hand, size of the aorta wall and
its nonlinear response to possible strain should also be considered. If a stent
is too large for the aorta, it is conceivable that in addition to flow there
will be external mechanical force from the stent acting on the aortic wall.
Thus a stent with diameter 12 mm should be the optimal solution for the current
patient-specific aorta, which agrees with the physicians' independent choice in
this case.

\section{Discussion and Conclusion}

Image-based in silico stent implantation
\cite{Pionteck2021,Chen2018,Kan2021,Neugebauer2016} and CFD
\cite{Zhong2016,Berg2018} together provide a new framework for stent planning
and interventional procedure evaluation. Besides a protocol for virtual
geometry deformation, a CFD method is needed to accurately resolve the flows in
the aortic geometries. However, blood flow in patient-specific aorta is
complicated \cite{Stein1976, Ku1997}. Laminar flow, turbulent flow and
transition between them may coexist spatiotemporally. In our study, we firstly
evaluated the accuracy of LES to predict such complicated flow. Two CFD
methods, the LBM based LES and DNS, in cooperation with flow MRI, were
considered. Both phantom and in vivo validations show that the LBM based LES,
which keeps a balance between numerical accuracy and computational requirement,
is a reasonable choice for resolving aortic flow. The validated LES was then
used to predict the flows in virtually deformed geometries with different stent
diameters. By comparing the flow fields, pressure drop, and maximum WSS, it was
found that the optimal stent was the one with diameter 12 mm, which agrees with
the physicians' independent choice.   

To restore blood flow, in addition to numerical methods, accurate geometry and
boundary conditions are also important. Based on MRI scans, aortic geometry can
be segmented and reconstructed from high-quality image slices
\cite{M.A.Hadhoud2012,Avendi2016,Kan2021}. Furthermore, flow MRI is used for
visualization and quantification of aortic flow. 2D flow MRI with through-plane
velocity encoding is usually performed in  clinical applications
\cite{Diagnosis2014,Tanaka2010,Wentland2013}. But the 2D flow measurement is
affected by the selection of the cross-sectional plane. 4D flow MRI,
alternatively, is able to obtain time-dependent 3D blood flow, which is
resolved in all three dimensions of space and the dimension of time during the
cardiac cycle. It can used for the estimation of the flow pathways and the WSS.
But 4D flow imaging takes a significant amount of time, which prevents wide
clinical application. Thus for patient-specific in silico stent implantation,
MRI and LES need to work together and both are indispensable. Particularly, MRI
provides data for geometry and boundary conditions, while LES predicts aortic
flows for further evaluation. 

In this work, we validated the LBM based LES with both phantom and in vivo
experiments, and provided a realistic example of the in silico stent implantation.
There are still many improvements possible which could be considered in the
future. Firstly, the aortic flow is unsteady, but we didn't consider time
dependent boundary condition in our simulations. We argue that current boundary
conditions are enough for us to compare different methods and provide results
for stent evaluation. By modeling the realistic cardiac cycle, one may get more
instantaneous flow information at the cost of longer computation time.
Secondly, we assigned the flow rates to the vessel branches according to their
cross areas. An alternative is the Windkessel model which considers resistance
and capacitance of the vessel network. Thirdly, we used a fast geometric method
to mimic the complex interaction between the aortic wall and the stent.
Ideally, one would take into account the mechanical properties of the aortic
wall and the stent, and model their interaction using finite element method.
Unfortunately it is still a big challenge to get accurate orthotropic
properties of the fiber reinforced aortic wall and to model the contact problem
numerically. Thus a simplified geometric method is a reasonable start. Finally,
this methodology could also be extended to other stenosis, such as cerebral
artery stenosis.

\section*{Conflict of interest}
The authors have no conflict to disclose.
\section*{Acknowledgments}
Computational resources from HPC at GWDG and MPCDF are appreciated.
Scientific comments and suggestions from the reviewers are also gratefully acknowledged. 
\section*{Ethics Declarations}
This article does not describe studies with human or animal subjects. 
The anonymous images used in this article were obtained from
a clinical database. Informed consent of the patients that permits
research use of their data has been obtained beforehand.

\bibliographystyle{apalike}
\bibliography{references}

\end{document}